\documentclass[a4paper]{IEEEtran}
\usepackage{color}
\usepackage{leftidx}
\usepackage{cite}
\usepackage{microtype}
\usepackage{cleveref}
\usepackage{subfigure}
\usepackage{graphicx,amssymb,amstext,amsmath}
\usepackage[ruled,vlined,lined,ruled,linesnumbered]{algorithm2e}
\begin{document}

\title{An ADMM Based Method for Computation Rate Maximization in Wireless Powered Mobile-Edge Computing Networks}

\author{Suzhi~Bi$^*$ and Ying-Jun~Angela~Zhang$^\dagger$\\
$^*$College of Information Engineering, Shenzhen University, Shenzhen, Guangdong, China 518060\\
$^\dagger$Department of Information Engineering, The Chinese University of Hong Kong, Shatin, N.T., Hong Kong SAR\\
E-mail:~bsz@szu.edu.cn, yjzhang@ie.cuhk.edu.hk \\
\thanks{This work was supported in part by the National Natural Science Foundation of China under Project 61501303, the Foundation of Shenzhen City under Project JCYJ20160307153818306, and the Science and Technology Innovation Commission of Shenzhen under Project 827/000212. The work of Y-J.~A.~Zhang was supported in part by General Research Funding (Project number 14200315) from the Research Grants Council of Hong Kong and Theme-Based Research Scheme (Project number T23-407/13-N). }\vspace{-2ex}}

\maketitle

\begin{abstract}
In this paper, we consider a wireless powered mobile edge computing (MEC) network, where the distributed energy-harvesting wireless devices (WDs) are powered by means of radio frequency (RF) wireless power transfer (WPT). In particular, the WDs follow a binary computation offloading policy, i.e., data set of a computing task has to be executed as a whole either locally or remotely at the MEC server via task offloading. We are interested in maximizing the (weighted) sum computation rate of all the WDs in the network by jointly optimizing the individual computing mode selection (i.e., local computing or offloading) and the system transmission time allocation (on WPT and task offloading). The major difficulty lies in the combinatorial nature of multi-user computing mode selection and its strong coupling with transmission time allocation. To tackle this problem, we propose a joint optimization method based on the ADMM (alternating direction method of multipliers) decomposition technique. Simulation results show that the proposed method can efficiently achieve near-optimal performance under various network setups, and significantly outperform the other representative benchmark methods considered. Besides, using both theoretical analysis and numerical study, we show that the proposed method enjoys low computational complexity against the increase of networks size.
\end{abstract}
\vspace{-2ex}

\IEEEpeerreviewmaketitle

\section{Introduction}
Finite battery lifetime and low computing capability of size-constrained wireless devices (WDs) have been longstanding performance limitations of many low-power wireless networks, e.g., wireless sensor networks (WSNs) and Internet of Things (IoT), especially for supporting many emerging applications that require sustainable and high-performance computations, e.g., autonomous driving and augmented reality.

Radio frequency (RF) based \emph{wireless power transfer} (WPT) has been recently identified as an effective solution to the finite battery capacity problem \cite{2015:Bi,2016:Bi,2016:Bi1,2016:Bi2}. Specifically, WPT uses dedicated RF energy transmitter, which can continuously charge the battery of remote energy-harvesting devices. Thanks to the broadcasting nature of RF signal, WPT is particularly suitable for powering a large number of closely-located WDs, like those deployed in WSNs and IoT. On the other hand, a recent technology innovation named \emph{mobile edge computing} (MEC) has attracted massive industrial investment and has been identified as a key technology towards future 5G network \cite{2016:Chiang,2017:Mao,2015:ETSI}. As its name suggests, MEC allows the WDs to offload intensive computations to nearby servers located at the edge of radio access network, e.g., cellular base station and WiFi access point (AP), to reduce computation latency and energy consumption. In general, there are two basic computation task offloading models in MEC, i.e., binary and partial computation offloading \cite{2017:Mao}. Specifically, \emph{binary offloading} requires a task to be executed as a whole either locally at the WD or remotely at the MEC server. Partial offloading, on the other hand, allows a task to be partitioned into two parts with one executed locally and the other offloaded for edge execution. In practice, binary offloading is suitable for simple tasks that are not partitionable, while partial offloading is favorable for some complex tasks composed of multiple parallel segments. A key research problem is the joint design of task offloading and system resource allocation to optimize the computing performance, which has been extensively studied under both binary and partial computation offloading policies \cite{2013:Wu,2016:Wang,2017:You,2016:Chen}.

The integration of WPT and MEC technologies introduces a new paradigm named \emph{wireless powered MEC}, where the distributed MEC wireless devices are powered by means of WPT. The deployment of wireless powered MEC systems can potentially tackle the two aforementioned performance limitations in low-power wireless networks like IoT. Compared to conventional battery-powered MEC, the optimal design in a wireless powered MEC network is more challenging. On one hand, the task offloading and resource allocation decisions now depend on the distinct amount of energy harvested by individual WDs from WPT. On the other hand, WPT and task offloading need to share the limited wireless resource, e.g., time or frequency. There are few existing studies on wireless powered MEC system \cite{2016:You,2017:Wang1,2017:Wang2}. \cite{2016:You} considers a single-user wireless powered MEC with binary offloading, where the user maximizes its probability of successful computation under latency constraint. In a multi-user scenario, \cite{2017:Wang1} considers using a multi-antenna AP to power the users and minimizes the AP's total energy consumption. \cite{2017:Wang2} also considers maximizing the weighted sum computation rate of a multi-user wireless powered MEC network. However, both \cite{2017:Wang1} and \cite{2017:Wang2} assume partial computation offloading policy. Mathematically speaking, partial offloading is a convex-relaxed version of the binary offloading policy. In a multi-user environment, the optimal design under the binary offloading policy often involves non-convex combinatorial optimization problems, which is much more challenging and currently lacking of study.

In this paper, we consider a wireless powered MEC network as shown in Fig.~\ref{101}, where the AP is reused as both energy transmitter and MEC server that transfers RF power to and receives computation offload from the WDs. Each device follows the \emph{binary offloading policy}. In particular, we are interested in maximizing the \emph{weighted sum computation rate}, i.e., the number of processed bits per second, of all the WDs in the network. Our contributions are detailed below.
\begin{enumerate}
  \item We formulate a joint optimization of user computing mode selection and the system transmission time allocation. The combinatorial nature of multi-user computing mode selection makes the optimal solution hard to obtain in general. As a performance benchmark, an enumeration-based optimal method is presented for evaluating the proposed reduced-complexity algorithm.
  \item We devise an ADMM-based technique that tackles the hard combinatorial mode selection by decomposing the original problem into parallel small-scale integer programming subproblems, one for each WD. We further show that the computational complexity of the proposed method increases slowly at a linear rate $O(N)$ of the network size $N$.
  \item Extensive simulations show that both proposed algorithm can achieve \emph{near-optimal} performance under various network setups, and significantly outperform the other benchmark algorithms. Because of its $O(N)$ computational complexity, the proposed method is especially applicable to large-size IoT networks.
\end{enumerate}

\section{System Model}
\subsection{Network Model}
\begin{figure}
\centering
  \begin{center}
    \includegraphics[width=0.45\textwidth]{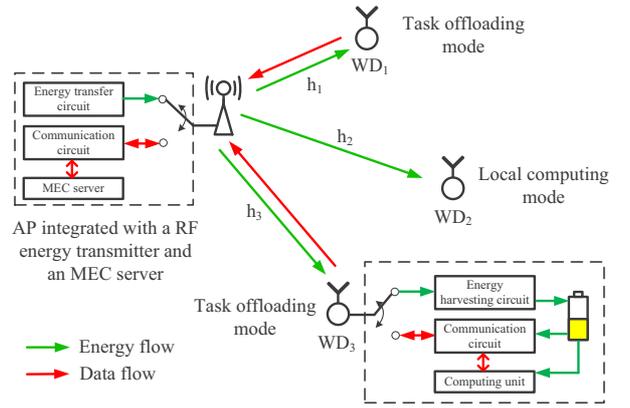}
  \end{center}
  \caption{An example $3$-user wireless powered MEC system with binary computation offloading.}
  \label{101}
\end{figure}

As shown in Fig.~\ref{101}, we consider a wireless powered MEC network consisting of an AP and $N$ WDs, where the AP and the WDs have a single antenna each. In particular, an RF energy transmitter and a MEC server is integrated at the AP. The AP is assumed to be connected to a stable power supply and broadcast RF energy to the distributed WDs, while each WD has an energy harvesting circuit and a rechargeable battery that can store the harvested energy to power its operations. Each device, including the AP and the WDs, has a communication circuit. Specifically, we assume that WPT and communication are performed in the same frequency band. To avoid mutual interference, the communication and energy harvesting circuits of each WD operate in a time-division-multiplexing (TDD) manner. A similar TDD circuit structure is also applied at the AP to separate energy transmission and communication with the WDs. Within each system time frame of duration $T$, the wireless channel gain between the AP and the $i$-th WD is denoted by $h_i$, which is assumed reciprocal for the downlink and uplink,\footnote{The channel reciprocity assumption is made to obtain more design insights on the impact of wireless channel. The proposed algorithm in this paper, however, can be easily extended to the case with non-equal uplink and downlink channels.} and static within each time frame but may vary across different time frames.

Within each time frame, we assume that each WD needs to accomplish a certain computing task based on its local data. For instance, a WD as a wireless sensor needs to regularly generate an estimate, e.g., the pollution level of the monitored area, based on the raw data samples measured from the environment. In particular, the computing task of a WD can be performed locally by the on-chip micro-processor, which has low computing capability due to the energy- and size-constrained computing processor. Alternatively, the WD can also offload the data to the MEC server with much more powerful processing power, which will compute the task and send the result back to the WD.

In this paper, we assume that the WDs adopt a binary computation offloading rule. That is, a WD must choose to operate in either the local computing mode (mode $0$, like WD$_2$ in Fig.~1) or the offloading mode (mode $1$, like WD$_1$ and WD$_3$) in each time frame. In practice, this corresponds to a wide variety of applications. For instance, the measurement samples of a sensor are correlated in time, and thus need to be jointly processed to enhance the estimation accuracy.

\subsection{Computation Model}
\begin{figure}
\centering
  \begin{center}
    \includegraphics[width=0.45\textwidth]{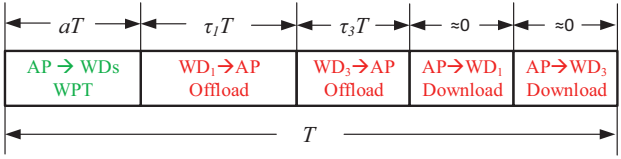}
  \end{center}
  \caption{An example time allocation in the $3$-user wireless powered MEC network in Fig.~\ref{101}. Only WD$_1$ and WD$_3$ selecting mode $1$ offload the task to and download the computation results from the AP.}
  \label{110}
\end{figure}

We consider an example transmission time allocation in Fig.~\ref{110}. We use two non-overlapping sets  $\mathcal{M}_0$ and $\mathcal{M}_1$ to denote the indices of WDs that operate in mode $0$ and $1$, respectively. As such $\mathcal{M} = \mathcal{M}_0 \cup \mathcal{M}_1 = \{1,\cdots,N\}$ is the set of all the WDs. In the first part of a tagged time frame, the AP broadcasts wireless energy to the WDs for $aT$ amount of time, where $a\in[0,1]$, and all the WDs harvest the energy. Specifically, the energy harvested by the $i$-th WD is
\begin{equation}
\label{103}
E_i = \mu P h_i a T,\ i=1,\cdots,N,
\end{equation}
where $P$ denotes the RF energy transmit power of the AP and $\mu\in(0,1)$ denotes the energy harvesting efficiency \cite{2015:Bi}. In the second part of the time frame $(1-a)T$, the WDs in $\mathcal{M}_1$ (e.g., WD$_1$ and WD$_3$ in Fig.~\ref{101}) offload the data to the AP. To avoid co-channel interference, we assume that the WDs take turns to transmit in the uplink, and the time that a WD$_i$ transmits is denoted by $\tau_i T$, $\tau_i\in [0,1]$. Depending on the selected computing mode, the detailed operation of each WD is illustrated as follows.

\subsubsection{Local Computing Mode}
Notice that the energy harvesting circuit and the computing unit are separate. Thus, a mode-$0$ WD can \emph{harvest energy and compute its task simultaneously}. Let $\phi>0$ denote the number of computation cycles needed to process one bit of raw data, which is assumed equal for all the WDs. Let $f_i$ denote the processor's chosen computing speed (cycles per second) and $0\leq t_i\leq T$ denote the computation time of the WD. The power consumption of the processor is modeled as $k_i f_i^3$ (joule per second), where $k_i$ denotes the computation energy efficiency coefficient of the processor's chip \cite{2016:Wang}. Then, the total energy consumption is constrained by
\begin{equation}
\label{2}
k_i f_i^3 t_i \leq E_i
\end{equation}
to ensure sustainable operation of the WD.\footnote{We assume each WD has sufficient initial energy in the very beginning and the battery capacity is sufficiently large such that battery-overcharging is negligible. Besides, for simplicity, we do not assume a maximum computing speed for the WDs considering their low harvested energy.} With the above computation model, the computation rate of WD$_i$ (in bits per second) denoted by $r_i$, can be calculated as \cite{2016:Wang}
\begin{equation}
\label{1}
r_i = \frac{f_i t_i}{\phi T}\leq \frac{1}{\phi T}\left(\frac{E_i}{k_i}\right)^{\frac{1}{3}}t_i^{\frac{2}{3}},\ \forall i \in \mathcal{M}_0,
\end{equation}
where the inequality is obtained from (\ref{2}). Therefore, the maximum $r_i^*$ is achieved by setting $t_i^*=T$, i.e., the WD computes for a maximal allowable time throughout the time frame and at a minimal possible computing speed. By substituting $t_i^*=T$ and $f_i^*=\left(\frac{E_i}{k_i T}\right)^{\frac{1}{3}}$ into (\ref{1}), the maximum local computation rate of a mode-$0$ WD is
\begin{equation}
\label{114}
r_i^*  = \frac{f_i^* t_i^*}{\phi T} =\eta_1 \left(\frac{h_i}{k_i}\right)^{\frac{1}{3}} a^{\frac{1}{3}},\ \forall i\in \mathcal{M}_0,
\end{equation}
where $\eta_1 \triangleq  \frac{\left( \mu P \right)^{\frac{1}{3}}}{\phi}$ is a fixed parameter.

\subsubsection{Offloading Mode}
Due to the TDD circuit constraint, a mode-$1$ WD can only \emph{offload its task to the AP after harvesting energy}. We denote the number of bits to be offloaded to the AP as $v_u b_i$, where $b_i$ denotes the amount of raw data and $v_u>1$ indicates the communication overhead in task offloading, such as packet header and encryption. Let $P_i$ denote the transmit power of the $i$-th WD. Then, the maximum $b_i^*$ equals to the data transmission capacity, i.e.,
\begin{equation}
\label{3}
b_i^* =  \frac{B\tau_i T}{v_u}\log_2\left(1+\frac{P_i h_i}{ N_0}\right), \ \forall i\in \mathcal{M}_1,
\end{equation}
where $B$ denotes the communication bandwidth and $N_0$ denotes the receiver noise power.

After receiving the raw data of all the WDs, the AP computes and sends back the output result of length $r_d b_i$ bits back to the corresponding WD. Here, $r_d \ll 1$ indicates the output/input ratio including the overhead in downlink transmission. In practice, the computing capability and the transmit power of the AP is much stronger than the energy-harvesting WDs, e.g., by more than three orders of magnitude. Beside, $r_d$ is a very small value, e.g., one output temperature estimation from tens of input sensing sample. Accordingly, we neglect the time spent on task computation and feedback by the AP like in \cite{2013:Wu,2016:You,2017:Wang1}. In this case, task offloading can occupy the rest of the time frame after WPT, i.e., $\sum_{i\in \mathcal{M}_1} \tau_i + a \leq 1$. Besides, from the above discussion, we also neglect the energy consumption by the WD on receiving the result from the AP and consider only the energy consumptions on data transmission to the AP. In this case, the WD should exhaust its harvested energy on task offloading, i.e., $P_i^* = E_i/\tau_i T$, to maximize its computation rate. From (\ref{3}), the maximum computation rate of a mode-$1$ WD$_i$ is
\begin{equation}
\label{4}
r_i^* =\frac{b_i^*}{T} =  \frac{B\tau_i}{v_u}\log_2\left(1+\frac{\mu P a h_i^2}{ \tau_i N_0}\right), \ \forall i\in \mathcal{M}_1.
\end{equation}

\subsection{Problem Formulation}
In this paper, we maximize the weighted sum computation rate of all the WDs in each time frame. From (\ref{114}) and (\ref{4}), the computation rates of the WDs are related to their computing mode selection and the system resource allocation on WPT and task offloading. Mathematically, the computation rate maximization problem is formulated as follows.
 \begin{subequations}
   \begin{align}
  (P1): \ & \underset{\mathcal{M}_0,a, \boldsymbol{\tau}}{\text{maximize}} & &  \sum_{i \in \mathcal{M}_0} w_i  \eta_1 \left(\frac{h_i}{k_i}\right)^{\frac{1}{3}} a^{\frac{1}{3}} \\
    & & & +  \sum_{j \in \mathcal{M}_1}  w_j \varepsilon \tau_j\ln \left(1+ \frac{\eta_2 h_j^2 a}{\tau_j}\right) \\
    & \text{subject to} &  & \sum_{j \in \mathcal{M}_1} \tau_j + a \leq 1, \label{116}\\
    & & & a \geq 0,\ \tau_j\geq 0, \ \forall j\in \mathcal{M}_1,\\
    & & & \mathcal{M}_0  \subseteq \mathcal{M}, \ \mathcal{M}_1 = \mathcal{M}\setminus  \mathcal{M}_0.
   \end{align}
\end{subequations}
Here, $\eta_2 \triangleq \frac{\mu P}{N_0}$ and $\varepsilon \triangleq \frac{B}{v_u \ln 2}$. $w_i>0$ denotes the weight of the $i$-th WD. $\boldsymbol{\tau}= \left\{\tau_j| j\in\mathcal{M}_1\right\}$ denotes the offloading time of the mode-$1$ WDs. The two terms of the objective function correspond to the computation rates of mode-$0$ and mode-$1$ WDs, respectively. (\ref{116}) is the time allocation constraint.

Due to the stringent energy and computation limitations of the WDs, we adopt a centralized control scheme where the AP is responsible for all the computations and coordinations, including selecting the computing mode for each WD. Among all the parameters in (P1), the AP only needs to estimate the wireless channel gains $h_i$'s that are time varying in each time frame. The others are static parameters that remain constant for sufficiently long period of time, such as $w_i$'s and $k_i$'s. Then, the AP calculates (P1) and broadcasts the solution $\{\mathcal{M}_0^*, a^*,\boldsymbol{\tau}^*\}$ to the WDs, which will react by operating in their designated computing modes.\footnote{The energy and time consumed on channel estimation and coordination can be modeled as two constant terms that will not affect the validity of the proposed algorithm. They are neglected in this paper for simplicity.}

Problem (P1) is a hard non-convex problem due to the combinatorial computing mode selection. However, we observe that the second term in the objective is jointly concave in $(a,\tau_j)$. Once $\mathcal{M}_0$ is given, (P1) reduces to a convex problem, where the optimal time allocation $\left\{a^*,\boldsymbol{\tau}^*\right\}$ can be efficiently solved using off-the-shelf optimization algorithms, e.g., interior point method \cite{2004:Boyd}. Accordingly, a straightforward method is to enumerate all the $2^N$ possible $\mathcal{M}_0$ and output the one that yields the highest objective value. The enumeration-based method may be applicable for a small number of WDs, e.g., $N\leq 10$, but quickly becomes computationally infeasible as $N$ further increases. Therefore, it will be mainly used as a benchmark to evaluate the performance of the proposed reduced-complexity algorithm in this paper. Before entering formal discussions on the algorithm design, it is worth mentioning that a closely related max-min rate optimization problem, which maximizes the minimum computation rate among the WDs, has its dual problem in the form of weight-sum-rate-maximization like (P1). In this sense, the proposed method in this paper can also be extended to enhance the \emph{user fairness} performance.

\section{An ADMM-Based Joint Optimization Method}\label{joint}

\subsection{Reformulation of (P1)}
In this section, we propose an ADMM-based method to solve (P1). The main idea is to decompose the hard combinatorial optimization (P1) into $N$ parallel smaller integer programming problems, one for each WD. Conventional decomposition techniques, such as dual decomposition, cannot be directly applied to (P1) due to the coupling factors in both objective and constraint. We first reformulate (P1) as an equivalent integer programming problem by introducing binary decision variables $m_i$'s and additional artificial variables $x_i$'s and $z_i$'s as follows
 \begin{equation}
\label{12}
   \begin{aligned}
    & \underset{a, \mathbf{z},\mathbf{x},\boldsymbol{\tau},\mathbf{m}}{\text{maximize}} & &  \sum_{i=1}^N w_i \bigg\{ \left(1-m_i\right)\eta_1 \left(\frac{h_i}{k_i}\right)^{\frac{1}{3}} x_i^{\frac{1}{3}}  \\
    & & & + m_i \varepsilon \tau_i \ln\left(1+ \frac{\eta_2 h_i^2 x_i}{\tau_i}\right)\bigg\}\\
    & \text{subject to} &  & \sum_{i=1}^N z_i + a \leq 1,\\
    &  & & x_i = a, z_i = \tau_i \ i=1,\cdots,N, \\
    &  & &   a, z_i, x_i,\tau_i\geq 0,\ m_i\in\left\{0,1\right\}, \ i=1,\cdots,N.
   \end{aligned}
\end{equation}
Here, $m_i=0$ for all $i\in \mathcal{M}_0$ and $m_i=1$ for all $i\in\mathcal{M}_1$. $\mathbf{z}=[z_1,\cdots,z_N]'$ and $\mathbf{x} = [x_1,\cdots,x_N]'$. With a bit abuse of notation, we denote $\boldsymbol{\tau} = [\tau_1,\cdots,\tau_N]'$. Notice that variables $z_i$ and $\tau_i$ are immaterial to the objective if $m_i = 0$. Then, (\ref{12}) can be equivalently written as
\begin{subequations}
\label{13}
   \begin{align}
    & \underset{a, \mathbf{z},\mathbf{x},\boldsymbol{\tau},\mathbf{m}}{\text{maximize}} & &  \sum_{i=1}^N  q_i(x_i,\tau_i,m_i) + g(\mathbf{z},a)\\
    & \text{subject to} &  & x_i = a, \tau_i = z_i \ i=1,\cdots,N, \label{122}\\
    &  & &   x_i,\tau_i\geq 0,\ m_i\in\left\{0,1\right\}, \ i=1,\cdots,N,
   \end{align}
\end{subequations}
where
\begin{equation*}
\begin{aligned}
&q_i(x_i,\tau_i,m_i) \\
=& w_i \left\{ \left(1-m_i\right)\eta_1 \left(\frac{h_i}{k_i}\right)^{\frac{1}{3}} x_i^{\frac{1}{3}} +  m_i \varepsilon \tau_i \ln\left(1+ \frac{\eta_2 h_i^2 x_i}{\tau_i}\right)\right\},
\end{aligned}
\end{equation*}
and
\begin{equation}
\label{42}
g(\mathbf{z},a) = \begin{cases}
0,&   \text{if} \left(\mathbf{z},a\right) \in \mathcal{G},\\
-\infty,   &   \text{otherwise},\\
\end{cases}
\end{equation}
where
\begin{equation*}
\mathcal{G} = \left\{\left(\mathbf{z},a\right)\mid \sum_{i=1}^N z_i +a \leq 1, a \geq 0, z_i \geq 0, i=1,\cdots,N \right\}.
\end{equation*}

Problem (\ref{13}) can be effectively decomposed using the ADMM technique \cite{2011:Boyd}, which solves for the optimal dual soulution. By introducing multipliers to the constraints in (\ref{122}), we can write a partial augmented Lagrangian of (\ref{13}) as
\begin{equation*}
\begin{aligned}
&L\left(\mathbf{u},\mathbf{v},\boldsymbol{ \theta}\right) =  \sum_{i=1}^N  q_i(\mathbf{u}) + g(\mathbf{v}) + \sum_{i=1}^N \beta_i \left(x_i - a\right) \\
&+ \sum_{i=1}^N \gamma_i \left(\tau_i - z_i\right)  - \frac{c}{2} \sum_{i=1}^N \left(x_i - a\right)^2 - \frac{c}{2} \sum_{i=1}^N \left(\tau_i - z_i\right)^2,
\end{aligned}
\end{equation*}
where $\mathbf{u} = \left\{\mathbf{x},\boldsymbol{\tau},\mathbf{m}\right\}$, $\mathbf{v}=\left\{\mathbf{z},a\right\}$, and $\boldsymbol{\theta} = \{\boldsymbol{\beta},\boldsymbol{\gamma}\}$. $c>0$ is a fixed step size. The corresponding dual function is
\begin{equation*}
d(\boldsymbol{\theta}) = \underset{\mathbf{u},\mathbf{v}}{\text{maximize}}\ \left\{L\left(\mathbf{u},\mathbf{v},\boldsymbol{ \theta}\right)\mid \mathbf{x}\geq \mathbf{0},\boldsymbol{\tau}\geq \mathbf{0},\mathbf{m}\in \mathbb{B}^{N\times 1}\right\},
\end{equation*}
where $\mathbb{B}^{N\times 1}$ denotes a $(N\times 1)$ binary vector. Furthermore, the dual problem is
\begin{equation}
\label{37}
\underset{\boldsymbol{\theta}}{\text{minimize}}\ d\left(\boldsymbol{\theta}\right).
\end{equation}

\subsection{Proposed ADMM Iterations}
The ADMM technique solves the dual problem (\ref{37}) by iteratively updating $\mathbf{u}$, $\mathbf{v}$, and $\boldsymbol{\theta}$. We denote the values in the $l$-th iteration as $\left\{\mathbf{u}^l, \mathbf{v}^l, \boldsymbol{\theta}^l\right\}$. Then, in the $(l+1)$-th iteration, the update of the variables is performed sequentially as follows:
\subsubsection{Step 1} Given $\left\{\mathbf{v}^l,\boldsymbol{\theta}^l\right\}$, we first maximize $L$ with respect to $\mathbf{u}$, where
\begin{equation}
\label{20}
\mathbf{u}^{l+1} = \arg \ \underset{\mathbf{u}}{\text{maximize}}\  L\left(\mathbf{u},\mathbf{v}^l,\boldsymbol{\theta}^l\right).
\end{equation}
Notice that (\ref{20}) can be decomposed into $N$ parallel subproblems. Each subproblem solves
\begin{equation}
\label{38}
\{x_i^{l+1},\tau_i^{l+1},m_i^{l+1}\}= \arg \underset{x_i, \tau_i \geq 0, m_i\in\{0,1\}}{\text{maximize}}\ s^l(x_i,\tau_i,m_i),
\end{equation}
where
\begin{equation*}
\begin{aligned}
&s_i^l(x_i,\tau_i,m_i) \\
=& q_i\left(x_i,\tau_i,m_i\right) + \beta_i^{l} x_i + \gamma_i^l \tau_i - \frac{c}{2}\left(x_i-a^l\right)^2 - \frac{c}{2}\left(\tau_i - z_i^l\right)^2.
\end{aligned}
\end{equation*}

We can equivalently express (\ref{38}) as
\begin{equation}
\label{16}
\begin{aligned}
\begin{cases}
\underset{x_i,\tau_i\geq 0}{\text{maximize}}\  &w_i \eta_1 \left(\frac{h_i}{k_i}\right)^{\frac{1}{3}} x_i^{\frac{1}{3}}+ \beta_i^{l} x_i + \gamma_i^l \tau_i - \frac{c}{2}\left(x_i-a^l\right)^2 \\
&- \frac{c}{2}\left(\tau_i - z_i^l\right)^2, \ \text{if } m_i=0,\\
\underset{x_i,\tau_i\geq 0}{\text{maximize}}\ & w_i \varepsilon \tau_i \ln \left(1+\frac{\eta_2 h_i^2 x_i}{ \tau_i}\right)+ \beta_i^{l} x_i + \gamma_i^l \tau_i \\
&- \frac{c}{2}\left(x_i-a^l\right)^2 - \frac{c}{2}\left(\tau_i - z_i^l\right)^2, \ \text{if } m_i=1.
\end{cases}
\end{aligned}
\end{equation}
For both $m_i=0$ and $1$, (\ref{16}) solves a strictly convex problem, and thus the optimal solution can be easily obtained, e.g., using the projected Newton's method \cite{2004:Boyd}. Accordingly, we can simply select $m_i=0$ or $1$ that yields a larger objective value in (\ref{16}) as $m_i^{l+1}$, and the corresponding optimal solution as $x_i^{l+1}$ and $\tau_i^{l+1}$. After solving the $N$ parallel subproblems, the optimal solution to (\ref{20}) is given by $\mathbf{u}^{l+1} = \left\{\mathbf{x}^{l+1},\boldsymbol{\tau}^{l+1},\mathbf{m}^{l+1}\right\}$. Notice that the complexity of solving each subproblem does not scale with $N$ (i.e., $O(1)$ complexity), thus the overall computational complexity of Step $1$ is $O(N)$.

\subsubsection{Step 2} Given $\mathbf{u}^{l+1}$, we then maximize $L$ with respect to $\mathbf{v}$. By the definition of $g(\mathbf{v})$ in (\ref{42}), $\mathbf{v}^{l+1} \in \mathcal{G}$ must hold at the optimum. Accordingly, the maximization problem can be equivalently written as the following convex problem
 \begin{equation}
 \label{43}
\begin{aligned}
&\mathbf{v}^{l+1} = \\
&\arg \underset{ \mathbf{z},a}{\text{maximize}} & & \sum_{i=1}^N \beta_i^l \left(x_i^{l+1} - a\right) + \sum_{i=1}^N \gamma_i^l \left(\tau_i^{l+1} - z_i\right) \\
 & & &- \frac{c}{2} \sum_{i=1}^N \left(x_i^{l+1} - a\right)^2 - \frac{c}{2} \sum_{i=1}^N \left(\tau_i^{l+1} - z_i\right)^2\\
 & \text{subject to} &  & \sum_{i=1}^N z_i +a \leq 1,\ a\geq 0, \ z_i\geq 0, i=1,\cdots,N.
\end{aligned}
 \end{equation}
Instead of using standard convex optimization algorithms, e.g., interior point method, here we devise an alternative low-complexity algorithm. By introducing a multiplier $\psi$ to the constraint $\sum_{i=1}^N z_i +a \leq 1$, it holds at the optimum that
\begin{equation}
\begin{aligned}
a^* &= \left(\frac{\sum_{i=1}^N x_i^{l+1}}{N} - \frac{\sum_{i=1}^N \beta_i^l + \psi^*}{cN}\right)^{+},\\
z_i^* &= \left(\tau_i^{l+1} - \frac{\gamma_i^l + \psi^*}{c}\right)^{+},\ i=1,\cdots,N,
\end{aligned}
\end{equation}
where $(x)^+ \triangleq \max\left(x,0\right)$. As $a^*$ and $z_i^*$ are non-increasing with $\psi^* \geq 0$, the optimal solution can be obtained by a bi-section search over $\psi^*\in (0,\bar{\psi})$, where $\bar{\psi}$ is a sufficiently large value, until $\sum_{i=1}^N z_i^* +a^* = 1$ is satisfied (if possible), and then comparing the result with the case of $\psi^* =0$ (the case that $\sum_{i=1}^N z_i^* +a^* < 1$). The details are omitted due to the page limit. Overall, the computational complexity of the bi-section search method to solve (\ref{43}) is $O(N)$.

\subsubsection{Step 3} Finally, given $\mathbf{u}^{l+1}$ and $\mathbf{v}^{l+1}$, we minimize $L$ with respect to $\boldsymbol{\theta}$, which is achieved by updating the multipliers $\boldsymbol{\theta}^l= \{\boldsymbol{\beta}^{l},\boldsymbol{\gamma}^l\}$ as
\begin{equation}
\label{22}
\begin{aligned}
\beta_i^{l+1} &= \beta_i^l - c(x_i^{l+1}-a^{l+1}),\ i=1,\cdots,N,\\
\gamma_i^{l+1} & = \gamma_i^l - c(\tau_i^{l+1}-z_i^{l+1}),\ i=1,\cdots,N.\\
\end{aligned}
\end{equation}
Evidently, the computational complexity of Step $3$ is $O(N)$.

The above Steps $1$ to $3$ repeat until a specified stopping criterion is met. In general, the stopping criterion is specified by two thresholds: absolute tolerance (e.g., $\sum_{i=1}^N |x_i^{l}-a^{l}|+ |\tau_i^l-z^l|$) and relative tolerance (e.g., $|a^{l}-a^{l-1}|+\sum_{i=1}^N |z_i^{l}-z_i^{l-1}|$) \cite{2011:Boyd}. The pseudo-code of the ADMM method solving (P1) is illustrated in Algorithm $1$. As the dual problem (\ref{37}) is convex in $\boldsymbol{\theta}= \{\boldsymbol{\beta},\boldsymbol{\gamma}\}$, the convergence of the proposed method is guaranteed. Meanwhile, the convergence of the ADMM method is insensitive to the choice of step size $c$ \cite{2015:Ghadimi}. Thus, we set $c = \varepsilon$ without loss of generality. Besides, we can infer that the computational complexity of one ADMM iteration (including the $3$ steps) is $O(N)$, because each of the $3$ steps has $O(N)$ complexity. Notice that the ADMM algorithm may not exactly converge to the primal optimal solution of (\ref{12}) due to the potential duality gap of non-convex problems. Therefore, upon termination of the algorithm, the dual optimal solution $\left\{a^l,\boldsymbol{\tau}^l,\mathbf{m}^l\right\}$ is an approximate solution to (\ref{12}), whose performance gap will be evaluated through simulations.

\begin{algorithm}
\small
 \SetAlgoLined
 \SetKwData{Left}{left}\SetKwData{This}{this}\SetKwData{Up}{up}
 \SetKwRepeat{doWhile}{do}{while}
 \SetKwFunction{Union}{Union}\SetKwFunction{FindCompress}{FindCompress}
 \SetKwInOut{Input}{input}\SetKwInOut{Output}{output}
 \Input{The number of WDs $N$ and other system parameters, e.g, $h_i$'s and $w_i$'s.}
 \textbf{initialization:}  $\{\boldsymbol{\beta}^0,\boldsymbol{\gamma}^0\}\leftarrow -100$;  $a^0 \leftarrow 0.9$;  $z_i^0 = (1-a^0)/N,\ i=1,\cdots, N$\;
 $c\leftarrow \varepsilon$, $\sigma_1\leftarrow 0.0005N$ , $l\leftarrow 0$\;
       \Repeat{$\sum_{i=1}^N \left(|x_i^{l}-a^{l}|+ |\tau_i^l-z^l|\right)<2\sigma_1$ \emph{and} $|a^{l}-a^{l-1}|+\sum_{i=1}^N |z_i^{l}-z_i^{l-1}|<\sigma_1$}{
      \For{\emph{each} WD$_i$}{
            Update local variables $\{x_i^{l+1},\tau_i^{l+1}, m_i^{l+1}\}$ by solving (\ref{16})\;
      }
      Update coupling variables $\left\{\mathbf{z}^{l+1},a^{l+1}\right\}$ by solving (\ref{43})\;

      Update multipliers $\{\boldsymbol{\beta}^{l+1},\boldsymbol{\gamma}^{l+1}\}$ using (\ref{22})\;
      $l\leftarrow l+1$\;
}
 \textbf{Return} $\left\{a^l,\boldsymbol{\tau}^l,\mathbf{m}^l\right\}$ as an approximate solution to (P1)\;
 \caption{ADMM-based joint mode selection and resource allocation algorithm}
\end{algorithm}

\section{Simulation Results}
In this section, we present simulations to evaluate the performance of the proposed algorithm. In all simulations, we use the parameters of the Powercast TX91501-3W transmitter with $P=3$W (Watt) as the energy transmitter at the AP, and those of P2110 Powerharvester as the energy receiver at each WD with $\mu= 0.51$ energy harvesting efficiency.\footnote{Please see the detailed product specifications on the website of Powercast Co. (http://www.powercastco.com).} Without loss of generality, we set $T=1$. The wireless channel gain $h_i$ follows the free-space path loss model $h_i = A_d\left(\frac{3\cdot10^8}{4\pi f_c d_i}\right)^{d_e},\ i=1,\cdots,N$, where $A_d = 4.11$ denotes the antenna gain, $f_c =915$ MHz denotes the carrier frequency, $d_i$ in meters denotes the distance between the WD$_i$ and AP, and $d_e\geq 2$ denotes the path loss exponent. Unless otherwise stated, $d_e=2.8$. Likewise, we set equal computing efficiency parameter $k_i = 10^{-26}$, $i=1,\cdots,N$, and $\phi=100$ for all the WDs \cite{2016:Wang}. For the data offloading mode, the bandwidth $B=2$ MHz, $v_u = 1.1$ and noise power $N_0=10^{-10}$ watt.

\subsection{Computation Rate Performance Comparisons}
We first evaluate the computation rate performance of the proposed ADMM-based algorithm. For performance comparisons, we consider the following three representative benchmark methods:
\begin{enumerate}
  \item Optimal: exhaustively enumerates all the $2^N$ combinations of $N$ WDs' computing modes;
  \item Offloading only: all the WDs offload their tasks to the AP, $\mathcal{M}_0 = \emptyset$;
  \item Local computing only: all the WDs perform computations locally, $\mathcal{M}_0 = \mathcal{M}$.
\end{enumerate}

\begin{figure}
  \centering
  \subfigure[Under different path loss exponent.]{\includegraphics[width=0.45\textwidth]{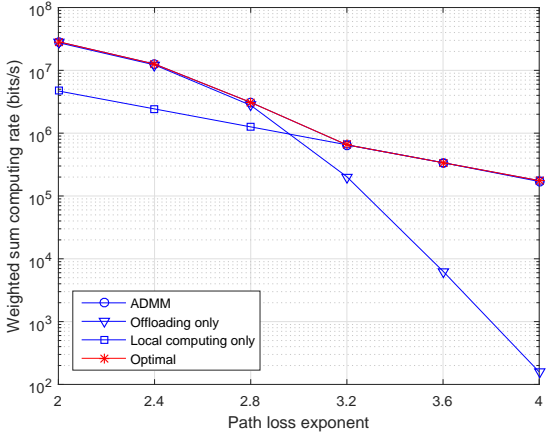}}\quad
  \subfigure[Under different average AP-WD distance.]{\includegraphics[width=0.45\textwidth]{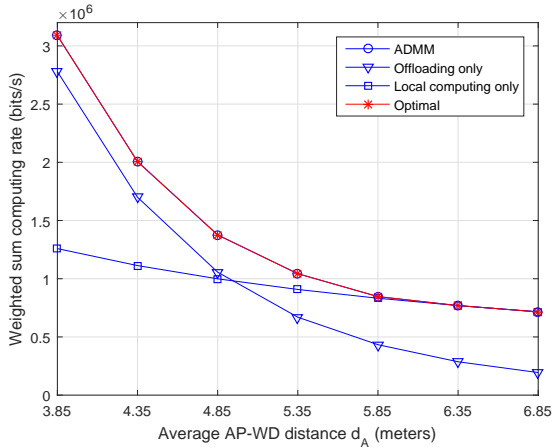}}
  \caption{Comparisons of computation rate performance of different algorithms when $N=10$. Figure above: when $d_e$ varies. Figure below: when $d_e=2.8$ and the average AP-to-WD distance varies.}
  \label{104}
\end{figure}

In Fig.~\ref{104}(a), we compare the weighted sum computation rate achieved by different schemes when the path loss exponent $d_e$ increases from $2$ to $4$. For the simplicity of illustration, we consider $N=10$ and set $d_i = 2.5 + 0.3(i-1)$ meters, $i=1\cdots,10$. In this case, the WDs are equally spaced by $0.3$ meter, where WD$_1$ ($d_1=2.5$) has the strongest wireless channel and WD$_{10}$ ($d_{10} = 5.2$) has the weakest wireless channel. Besides, we set $w_i=1$ if $i$ is an odd number and $w_i=2$ otherwise. We see that when $d_e$ is small and the wireless channels are strong, e.g., $d_e\leq 2.4$, the offloading-only scheme achieves near optimal solution. However, as we increase $d_e$, the performance of the offloading-only scheme quickly degrades, e.g., achieving only around $1/1000$ of the optimal rate when $d_e=4$, because the offloading rates severely suffer from the weak channels in both the uplink and downlink. In contrast, the local-computing-only scheme achieves the worst performance when $d_e$ is small (only around $1/6$ of the maximum when $d_e\leq 2.4$) but near-optimal performance when $d_e\geq 3.2$. On the other hand, the proposed ADMM method achieves near-optimal performance for all values of $d_e$ (at most $0.5\%$ performance gap compared to the optimal value).

In Fig.~\ref{104}(b), we fix $d_e = 2.8$ and compare the computation rate performance when the average distance $d_A$ between the AP and the WDs varies. For simplicity of illustration, we consider $10$ WDs uniformly placed within the range $[d_A-1.35,d_A+1.35]$ with a $0.3$ meter spacing between every two adjacent WDs. In this sense, the placement of the WDs in Fig.~\ref{104}(a) corresponds to $d_A = 3.85$. The weight assignment follows that in Fig.~\ref{104}(a). We observe that the proposed ADMM method achieves near-optimal performance for all values of $d_A$. The offloading-only scheme achieves relatively good performance when $d_A$ is small, e.g., $d_A\leq 4.35$, but poor performance when $d_A$ is large ($\approx \frac{1}{3}$ of the optimal value when $d_A=6.85$). The local-computing-only scheme, however, performs poorly when $d_A$ is small ($\approx \frac{1}{3}$ of the optimal value when $d_A=3.85$) but achieving near-optimal solution when $d_A$ is large. The results show that it is more preferable for a WD to offload computation when its wireless channel is strong and to perform local computing otherwise.

\begin{figure}
\centering
  \begin{center}
    \includegraphics[width=0.45\textwidth]{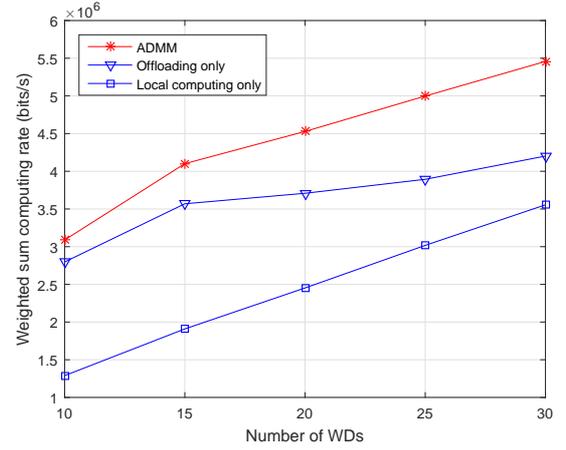}
  \end{center}
  \caption{Computation rate performance comparisons of different algorithms when the number of WDs $N$ varies.}
  \label{105}
\end{figure}

In Fig.~\ref{105}, we compare the performance of different algorithms when the number of WDs $N$ varies from $10$ to $30$. For each WD$_i$, its distance to the AP is uniformly generated as $d_i\thicksim U(2.5,5.2)$, and its weight $w_i$ is randomly assigned as either $1$ or $2$ with equal probability. Besides, each point in the figure is an average performance of $20$ independent random placements. Unlike in Fig.~\ref{104}, the optimal performance is not plotted because the mode-enumeration based optimal method is computationally infeasible for most values of $N$ within the considered range. For example, $N = 15$ needs to enumerate over $30000$ computing mode combinations. Instead, we only compare the performance of the other sub-optimal methods. We see that the proposed ADMM method significantly outperforms the other two benchmark methods, i.e., around $21\%$ and $92\%$ higher average computation rate than the offloading-only and local-computing-only schemes, respectively. In particular, the offloading-only scheme performs relatively well when $N\leq 15$, but the rate increase becomes slower than the other three methods when $N$ becomes larger.

To sum up from Fig.~\ref{104} and \ref{105}, the performance of the offloading-only and local-computing-only methods are very sensitive to the network parameters and placement, e.g., path loss exponent, distance, and network size, which may produce very poor performance in some practical setups. In contrast, even with fixed initial point, the proposed ADMM method can achieve near-optimal computation rate performance under different network setups.

\subsection{Computational Complexity Evaluation}
\begin{figure}
\centering
  \begin{center}
    \includegraphics[width=0.45\textwidth]{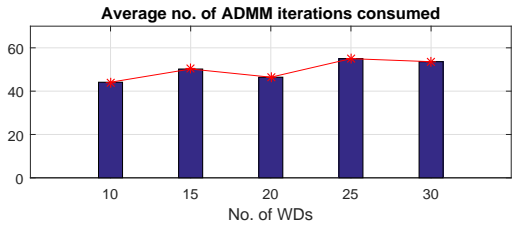}
  \end{center}
  \caption{Average number of iterations of Algorithm $1$ when the number of WDs varies.}
  \label{106}
\end{figure}

In Fig.~\ref{106}, we characterize the computational complexity of the proposed ADMM-based algorithm. Here, we use the same network setup as in Fig.~\ref{105} and plot the average number of iterations consumed by Algorithm $1$ before its convergence when the number of WDs varies. Interestingly, we observe that the ADMM-based method consumes almost constant number of iterations under different $N$ within the considered range, i.e., $O(1)$. As the computational complexities of one ADMM iteration is $O(N)$, the overall computational complexity of the ADMM-based method is $O(N)$ as well. The result indicates the complexity of the proposed ADMM based method increases slowly as the network size increase. Therefore, it is feasible to apply the ADMM-based method in a large-size IoT network where the network size dominates the overall complexity.

\section{Conclusions}
In this paper, we studied a weighted sum computation rate maximization problem in multi-user wireless powered edge computing networks with binary computation offloading policy. We formulated the problem as a joint optimization of individual computing mode selection and system transmission time allocation. In particular, we proposed an efficient ADMM-based method to tackle the hard combinatorial computing mode selection problem. Extensive simulation results showed that, with $O(N)$ time complexity, the proposed ADMM-based method can achieve near-optimal computation rate performance under different network setups, and significantly outperform the other representative benchmark methods.

\end{document}